\documentclass[aps,twocolumn,superscriptaddress,numerical]{revtex4-1}

\usepackage{amsmath}
\usepackage{epsfig}
\usepackage{graphicx}
\usepackage{bm}
\usepackage{amssymb}
\usepackage{slashed}
\usepackage{hyperref}
\hypersetup{
     colorlinks   = true,
     citecolor    = blue,
     urlcolor = blue,
     linkcolor = blue
}

\newcommand{\be}{\begin{equation}}
\newcommand{\ee}{\end{equation}}

\begin{document}

\title{Anomalous Nernst and Thermal Hall Effects in Tilted Weyl Semimetals}
\author{Yago Ferreiros}
\affiliation{Department of Physics, KTH Royal Institute of Technology, SE-106 91 Stockholm, Sweden}
\author{A. A. Zyuzin}
\affiliation{Department of Physics, KTH Royal Institute of Technology, SE-106 91 Stockholm, Sweden}
\affiliation{Ioffe Physical-Technical Institute, 194021 St. Petersburg, Russia}
\author{Jens H. Bardarson}
\affiliation{Department of Physics, KTH Royal Institute of Technology, SE-106 91 Stockholm, Sweden}
\affiliation{Max-Planck-Institut f\"ur Physik komplexer Systeme, 01187 Dresden, Germany}

\begin{abstract}
We study the anomalous Nernst and thermal Hall effects in a linearized low-energy model of a tilted Weyl semimetal, with two Weyl nodes separated in momentum space. For inversion symmetric tilt, we give analytic expressions in two opposite limits: for a small tilt, corresponding to a type-I Weyl semimetal, the Nernst conductivity is finite and independent of the Fermi level, while for a large tilt, corresponding to a type-II Weyl semimetal, it acquires a contribution depending logarithmically on the Fermi energy. This result is in a sharp contrast to the nontilted case, where the Nernst response is known to be zero in the linear model. The thermal Hall conductivity similarly acquires Fermi surface contributions, which add to the Fermi level independent, zero tilt result, and is suppressed as one over the tilt parameter at half filling in the Type-II phase. In the case of inversion breaking tilt, with the tilting vector of equal modulus in the two Weyl cones, all Fermi surface contributions to both anomalous responses cancel out, resulting in zero Nernst conductivity. We discuss two possible experimental setups, representing open and closed thermoelectric circuits. 

\end{abstract}

\maketitle

\section{Introduction}

Weyl semimetals are gapless topological materials whose band-structure consists of pairs of Weyl nodes separated in momentum space. First proposed to to occur in a class of iridate materials \cite{WTV11}, they were recently realized \cite{WFB15,HXBH15,H115,LXD15,LW15,SNY15,YLC15,XAH15,GB16}. 
In order for Weyl nodes to be located at different momenta, at least one of time reversal or inversion symmetry needs to be broken.
For a basic understanding of these materials at a theoretical level, the simplest realization consists of a single pair of Weyl nodes. 
Such a minimal model breaks time reversal \cite{BB11}, and if it additionally breaks inversion the Weyl nodes are also separated in energy. 
In contrast, for a time reversal symmetric Weyl semimetal, an inversion breaking model with a minimum of four Weyl points is needed \cite{HB12}.

Both time reversal and inversion breaking Weyl semimetals realize the chiral anomaly, which leads to a number of phenomena, such as the chiral magnetic effect \cite{FKW08,ZB12,CCG14} and the accompanying negative magnetoresistance \cite{NN83,HQ13,SS13,ZHJ15,XKO15,HZC15}, the chiral separation effect \cite{SZ04}, chiral magnetic waves \cite{KY11}, and chiral anomaly induced plasmon modes \cite{ZCX15}. 
A key feature that differentiates the time reversal breaking Weyl semimetal is the presence of a topological Chern-Simons term in its low-energy electromagnetic description.
This term gives rise to many of the predicted exotic physical properties of Weyl semimetals, such as a finite quantum anomalous Hall response \cite{ZB12}; optical properties such as birefringence \cite{CJ90,G12}, circular dichroism \cite{HQ15}, magnon electrodynamics \cite{HZR14}, helicons \cite{PKP15}, and the appearance of novel collective electromagnetic excitations \cite{FC16,KL16}; and a new mechanism for the phonon Hall viscosity \cite{CFLV15,CFLV16}.

In the context of high-energy physics, Weyl fermion Lagrangians are always assumed to be Lorentz invariant. 
In condensed matter physics, however, there is no reason for a real material to preserve Lorentz symmetry. 
Therefore, terms forbidden in the high-energy description of Weyl fermions may arise. 
In lowest order in momentum a Lorentz nonsymmetric term can be added to the conical dispersion relation: $\mathcal{E}=\pm v|\bm k|+\bm C\cdot\bm k$ \footnote{This term assumes isotropy, more general anisotropic terms could be considered also.}.
This term tilts the Weyl cone in the direction of the vector $\bm C$, such that for large enough tilts $|\bm C|\ge v$ a Lifshitz phase transition occurs, and the Fermi surface surrounding the Weyl node transforms into two Fermi surfaces of electrons and holes.
The realization of this type-II Weyl semimetal phase~\cite{SB15,XZZ15} has been reported in recent experiments~\cite{XAH16,FBW16}.

Type-II Weyl semimetals are expected to give rise to new physical phenomena while still conserving some of the key ingredients of type-I Weyl semimetals. 
In particular, since the topological nature of the Weyl nodes survives the Lifshitz transition, both the chiral anomaly and the Chern-Simons term remain. 
Experimental signature of the chiral anomaly in a supposed type-II Weyl semimetal has been reported \cite{LLC17}. 
The activation of the anomaly in the type-II phase was suggested to depend on the alignment of the magnetic field with the tilt direction \cite{SB15,YYY16,UB16}, though a recent semiclassical calculation obtained an anomaly induced positive longitudinal magneto-conductivity for all directions of the applied magnetic field \cite{SGTA16}. 
The topological Chern-Simons term in the time reversal breaking case \footnote{The tilting term in the Hamiltonian breaks time reversal by itself, so when we refer to the time reversal breaking tilted Weyl semimetal, we mean that it is time reversal breaking even at zero tilt.} and all the associated phenomena listed above should similarly remain. 
In particular, and of importance to our work, the quantum anomalous Hall effect was shown to be finite \cite{ZT16,COL16}, receiving Fermi level dependent contributions from the tilting term, even in the Type-I phase \cite{ZT16,SAP17,WS17}.

In addition to electric transport, thermal responses also carry signatures of the exotic physics of Weyl semimetals, such as the chiral anomaly \cite{L14,CCG14,L16}.
A number of experimental studies of thermal transport in Weyl semimetals have recently been reported \cite{HBC16,CLL16,WMF17,GLF17,SRF17}, of particular interest for the present work is the measurement of the Nernst effect in the Weyl semimetal NbP \cite{WMF17}. 
On the theoretical side, thermal transport was explored for nontilted Weyl \cite{K14,LLF14,SGT16,SA16,LDS16}, Dirac \cite{SMT16}, double Weyl \cite{CF16}, and multi-Weyl \cite{GMSS17} semimetals, whereas an extension to a model of a nontilted Weyl semimetal in the presence of axial magnetic fields was considered in \cite{LL17}. 
More specifically, in Ref.~\onlinecite{LLF14} it was shown that the anomalous Nernst current vanishes in a linear model of a time reversal breaking Weyl semimetal, while in Refs.~\onlinecite{SGT16,GMSS17} finite results were obtained in lattice models. 
This can be understood from the fact that, in the linearized model for the nontilted Weyl semimetal, the quantum anomalous Hall conductivity does not receive Fermi surface contributions, whereas this no longer holds when quadratic and higher corrections are considered. 
With the Nernst effect being a purely a Fermi surface effect, and the Nernst conductivity related to the Hall conductivity through the Mott rule, this explains its absence in the linear model.
In nontilted Weyl semimetals, the anomalous Nernst response comes entirely from higher order corrections to the low-energy dispersion relation.

Here, we study anomalous thermal transport \cite{L14,L16} in time-reversal-breaking tilted Weyl semimetals, focusing particularly on the Nernst and thermal Hall effects. 
We show that in the presence of a nonzero tilting term, Fermi surface contributions are significant already at the lowest order in momentum, and therefore the Nernst response is finite even in the linear model. 
We also obtain Fermi surface contributions to the thermal Hall effect, which add to the Fermi level independent, zero tilt result. 
Our results hold deep in both the type-I and type-II phases, where the validity of the linear model is guaranteed.

\section{Thermal transport}
In this section we review the relevant thermal transport formalism \cite{L64,SS77,CHR97,BR15} that we will then apply to tilted Weyl semimetals in the next section.
Our starting point is Luttinger's phenomenological transport equations for the electric and energy DC currents \cite{L64}
\begin{gather}
\bm J_{\textrm{tr}}=\bm L^{(1)}\left(-\bm\partial A_0-\frac{T}{e}\bm\partial\frac{\mu}{T}\right)+\bm L^{(2)}\left(-\frac{1}{c^2}\bm\partial\phi+T\bm\partial\frac{1}{T}\right),\label{eq transport 1}\\
\bm J_{\textrm{tr}}^\mathcal{E}=\bm L^{(3)}\left(-\bm\partial A_0-\frac{T}{e}\bm\partial\frac{\mu}{T}\right)+\bm L^{(4)}\left(-\frac{1}{c^2}\bm\partial\phi+T\bm\partial\frac{1}{T}\right).
\label{eq transport 2}
\end{gather}
The tensors $\bm L^{(i)}$ are the transport coefficients, $A_0$ and $\phi$ are the electromagnetic scalar field and the gravitational field respectively, $\mu$ and \textit{T} are the chemical potential and temperature, $e$ is the charge of the electron, and $c$ is the velocity of light. 
The energy current results from the combination of heat current $\bm J^Q$ and energy transported by the electric current; the heat current can therefore be written as 
\be
\bm J_{\textrm{tr}}^Q=\bm J_{\textrm{tr}}^\mathcal{E}-\frac{\mu}{e}\bm J_{\textrm{tr}}.
\label{eq heat current 1}
\ee

The transport coefficients can be obtained from the Kubo formula as a linear response to $\bm \partial A_0$ and $\bm \partial \phi$ ~\cite{L64}.
One has, however, to carefully distinguish between the transport currents $\bm J_{\textrm{tr}}$ relevant to Eqs.~\eqref{eq transport 1} and~\eqref{eq transport 2}, and the total local currents $\bm J$ described by the Kubo formula \cite{SS77,CHR97,BR15}.
While in time reversal invariant systems these coincide, breaking time reversal can lead to the appearance of magnetization electric and energy currents, which are circulating currents which average to zero and therefore do not carry any transport \cite{CHR97}.
Explicitly, in this case we have $\bm J=\bm J_{tr}+\bm J_{\textrm{mag}}$ and $\bm J^\mathcal{E}=\bm J_{tr}^\mathcal{E}+\bm J_{\textrm{mag}}^\mathcal{E}$.

Let us now fix the chemical potential and allow for gradients of temperature, and consider a sample disconnected from current leads such that no net electric current flows. 
In this situation, the transport electric current vanishes and an electric field is generated (the gradient of gravitational potential is set to zero). 
From the transport Eqs. (\ref{eq transport 1}) and (\ref{eq transport 2}), and using Eq. (\ref{eq heat current 1}) we have
\begin{gather}
\bm\partial A_0=-\bm S\bm\partial T\label{eq voltage},\\
\bm J^Q_{\textrm{tr}}=-\bm K\bm\partial T,
\end{gather}
where the thermopower $\bm S$ and thermal conductivity $\bm K$ are functions of the transport coefficients
\begin{gather}
\bm S=\frac{1}{T}\big[(\bm L^{(1)})^{-1}\bm L^{(2)}-\frac{\mu}{e}\big],\label{eq thermopower}\\
\bm K=\frac{1}{T}\big[\bm L^{(4)}-\bm L^{(3)}(\bm L^{(1)})^{-1}\bm L^{(2)}\big].
\label{eq thermal conductivity}
\end{gather}
For a Fermi liquid in the limit $k_B T \ll |\mu|$ ($k_B$ is the Boltzmann constant), the Mott rule and the Wiedemann-Franz law relate the thermopower and thermal conductivity, respectively, to the electric conductivity $\bm L^{(1)}$ as
\begin{gather}
\bm S=eLT\big[\bm L^{(1)}(T=0)\big]^{-1}\frac{d\bm L^{(1)}(T=0)}{d\mu},\label{eq Mott}\\
\bm K=LT\bm L^{(1)}(T=0),
\label{eq Wiedemann-Franz}
\end{gather}
where the Lorentz number $L=\pi^2k_B^2/3e^2$.
Note that the electric conductivity in Eqs.~\eqref{eq Mott} and~\eqref{eq Wiedemann-Franz} is evaluated at zero temperature, as both the Mott rule and the Wiedemann-Franz law are obtained as an expansion in $k_B T \ll |\mu|$;
furthermore, and importantly for our purposes, both the Mott rule and the Wiedemann-Franz law have been proven to hold in the quantum Hall state \cite{SS77}.

Let us now recall the Onsager relations, which relate the transport coefficients under the application of time reversal \cite{Ons1,Ons2}
\be
L^{(1,4)}_{ij}(\bm b)=L^{(1,4)}_{ji}(-\bm b),\quad L^{(2)}_{ij}(\bm b)=L^{(3)}_{ji}(-\bm b),
\ee
where $\bm b$ is a shorthand notation for all fields that are odd under time reversal. These relations can be inferred by inspection of the Kubo formulas and the invariance of the system under simultaneous reversal of time and $\bm b$.

Plugging Eqs.~\eqref{eq Mott} and~\eqref{eq Wiedemann-Franz} into Eqs.~\eqref{eq thermopower} and~\eqref{eq thermal conductivity}, respectively, and using the Onsager relations, we can write all transport coefficients as functions of the electric conductivity alone
\begin{gather}
\bm L^{(2)}=eLT^2\,\frac{d\bm L^{(1)}}{d\mu}+\frac{\mu}{e}\bm L^{(1)},\label{eq L2}\\
\bm L^{(4)}=\big(LT^2+\frac{\mu^2}{e^2}\big)\bm L^{(1)}+2\mu LT^2\,\frac{d\bm L^{(1)}}{d\mu}
\label{eq L4}
\end{gather}
where $\bm L^{(1)}$ is evaluated at $T=0$. In arriving to Eq. (\ref{eq L4}) we have assumed $L^{(2)}_{ij}(\bm b)=L^{(2)}_{ji}(-\bm b)$, which follows from Eq.~\eqref{eq L2} plus the Onsager relation for $\bm{L}^{(1)}$, and neglected terms of order higher than $T^2$. Substituting these coefficients in Eqs. (\ref{eq transport 1}) and (\ref{eq transport 2}), and using Eq. (\ref{eq heat current 1}) we get, setting $\bm \partial A_0=\bm \partial \phi=0$
\begin{gather}
\bm J_{\textrm{tr}}=-eLT\,\frac{d\bm L^{(1)}}{d\mu}\,\bm\partial T,\label{eq electric current}\\
\bm J_{\textrm{tr}}^Q=-LT\bm L^{(1)}\,\bm\partial T.
\label{eq heat current 2}
\end{gather}
These are the essential relations \cite{JM80,LLF14,CF16,SMT16} we will use in the next section.

\section{Anomalous Nernst and thermal Hall effects in a tilted Weyl semimetal}
Let us consider a minimal model of a time reversal breaking Weyl semimetal with two Weyl nodes of oposite chirality separated in momentum space, and  a tilting term. 
The linearized Hamiltonian around each Weyl node $s=\pm$ is
\be
H_{s}=\hbar C_{s}(k_z-sQ)+s\hbar v\bm\sigma\cdot(\bm k-sQ\bm e_z)
\label{eq Hamiltonian}
\ee
where $2Q$ is the distance between the Weyl points in momentum space along $\bm e_z$, $v$ is the Fermi velocity when $C_s= 0$, and $\bm \sigma$ is a vector composed of the three Pauli matrices. The type of Weyl point is defined by the tilt parameters $C_{s}$, such that the Weyl point is of type-I if $|C_{s}|<v$ and type-II if $|C_{s}|>v$.

This tilted time-reversal-breaking model is in the anomalous quantum Hall state, and therefore has a nonzero intrinsic anomalous Hall conductivity in the $x$-$y$ plane $L^{(1)}_{xy}=\sigma_{xy}$ \cite{ZT16,SAP17,WS17}. 
The Mott rule and Wiedemann-Franz law apply, and the thermal transport is given by Eqs.~\eqref{eq electric current} and~\eqref{eq heat current 2}. 
Writing the conductivity tensor in the form
\be
\bm L^{(1)}=\left(\begin{array}{ccc}\sigma_{xx}&\sigma_{xy}&0\\-\sigma_{xy}&\sigma_{yy}&0\\0&0&\sigma_{zz}\end{array}\right),
\ee
we can directly extract the anomalous Nernst and thermal Hall currents
\begin{gather}
J_{a,\textrm{tr}}=-\alpha_{xy}\,\epsilon_{ab}\partial_b T,\label{eq nernst current}\\
J_{a,\textrm{tr}}^Q=-K_{xy}\,\epsilon_{ab}\partial_b T.
\label{eq thermal hall current}
\end{gather}
with the Nernst and thermal Hall conductivities given by
\begin{gather}
\alpha_{xy}=eLT\frac{d\sigma_{xy}}{d\mu},\label{eq nernst conductivity}\\
K_{xy}=LT\sigma_{xy}.\label{eq thermal Hall conductivity}
\end{gather}
Both responses are obtained by computing the value of the Hall conductivity as a function of $\mu$ (see Appendix \ref{appendix} and Refs. \cite{ZT16,SAP17} for a computation of $\sigma_{xy}$). 

\begin{figure}
\includegraphics[scale=0.40]{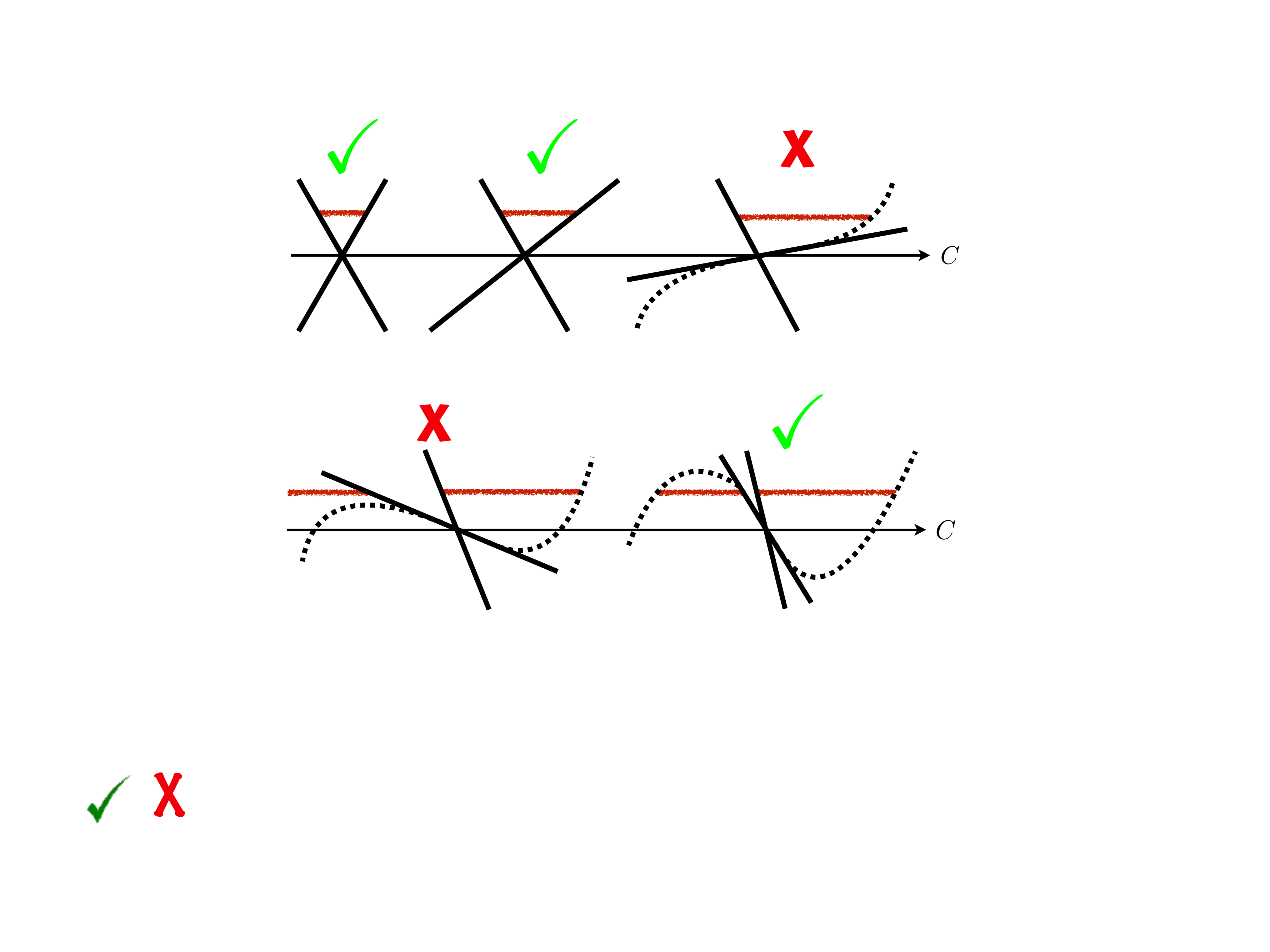}
\caption{Schematics of the competition between the linear and higher order momentum terms in the dispersion relation of a Weyl fermion. We show, from left to right, a Type-I (top) and a Type-II (bottom) Weyl cone for increasing tilt $C$. The dashed lines represent the deviation of the dispersion relation from the linear model due to higher order momentum corrections. The horizontal red lines represents the Fermi level. The check and X marks signal the validity and the failure of the linearized model, respectively.}
\label{fig. 2}
\end{figure}

We first consider the case of inversion symmetric tilt $C_+=-C_-=C$, where the coefficient $C$ can be either positive or negative. 
The linearized model is not well suited for computing the conductivity around the Lifshitz transition between type-I and type-II Weyl semimetals.
Starting in the type-I phase, by increasing the tilt the Fermi surfaces of each cone grows, reaching a point where quadratic terms in momentum compete with the linear terms (see Fig. \ref{fig. 2}(top)). 
For sufficiently small $\mu$, we can assume this to occur near the Lifshitz transition. 
The existence of this crossover at which higher order corrections become relevant means that the linear model is nonapplicable for values of the tilt near the transition. 
Further increasing the tilt, we go over into the type-II phase, at which point infinite electron and hole pockets would appear in pairs in the linear approximation; however, in a real material not only this pockets are of finite size, but some of them could even disappear, depending on the position of the Fermi level (Fig. \ref{fig. 2}(bottom-left)).
If we keep increasing the tilt, one expects that above a certain value of the tilt, and again assuming $\mu$ is sufficiently close to the Weyl nodes, electron and hole pockets will always coexist, in agreement with the linear model (Fig. \ref{fig. 2}(bottom-right)). 
For such values of the tilt deep in the type-II phase, higher order corrections in momentum are always important. 
A qualitatively correct description of the system can still be obtained with the linearized model by simply including a physical momentum cut-off, that accounts for the finiteness of the pockets due to nonlinearities. 
This momentum cut-off gives a measure of the Fermi surface size. 
Therefore, we will calculate the Nernst and thermal Hall conductivities deep in the type-I and type-II phases, excluding the regime $|C|\sim v$. 
In these two limits, analytic expressions can be written down.

\begin{figure*}
(a)
\begin{minipage}{.46\linewidth}
\includegraphics[scale=0.57]{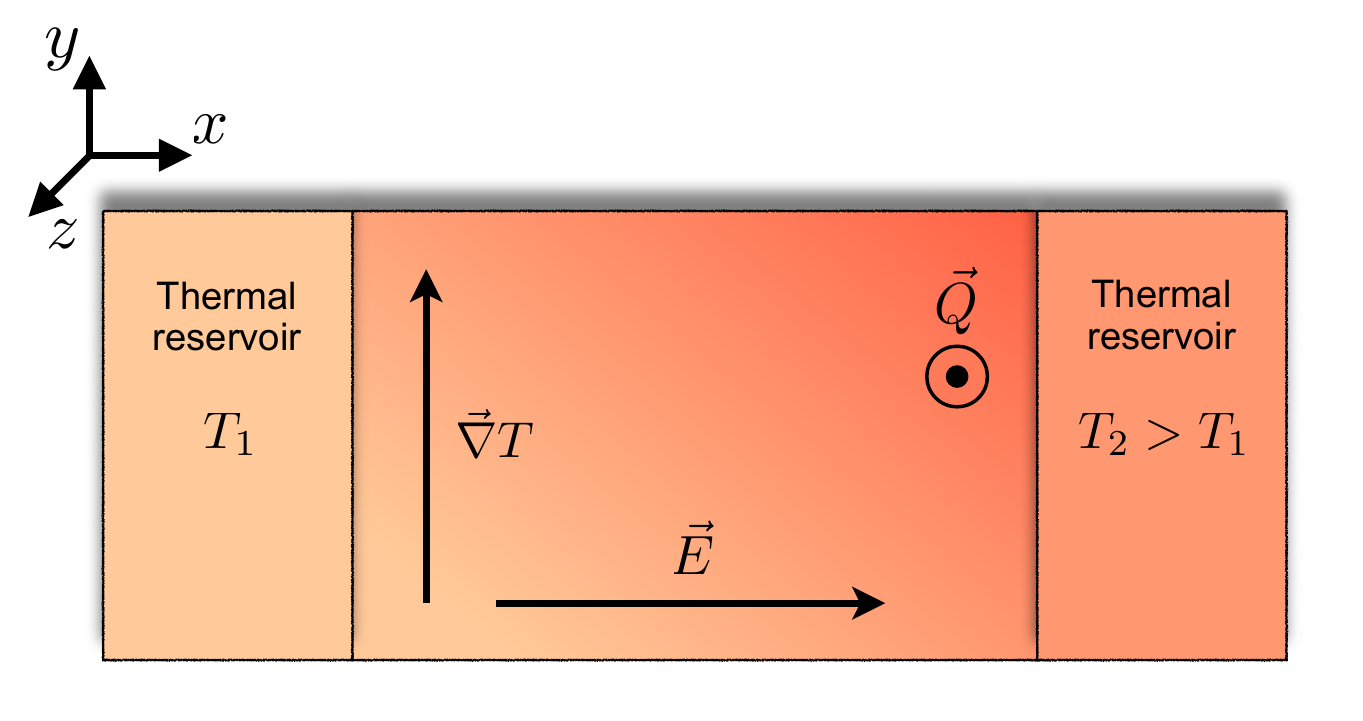}
\end{minipage}
(b)
\begin{minipage}{.46\linewidth}
\includegraphics[scale=0.57]{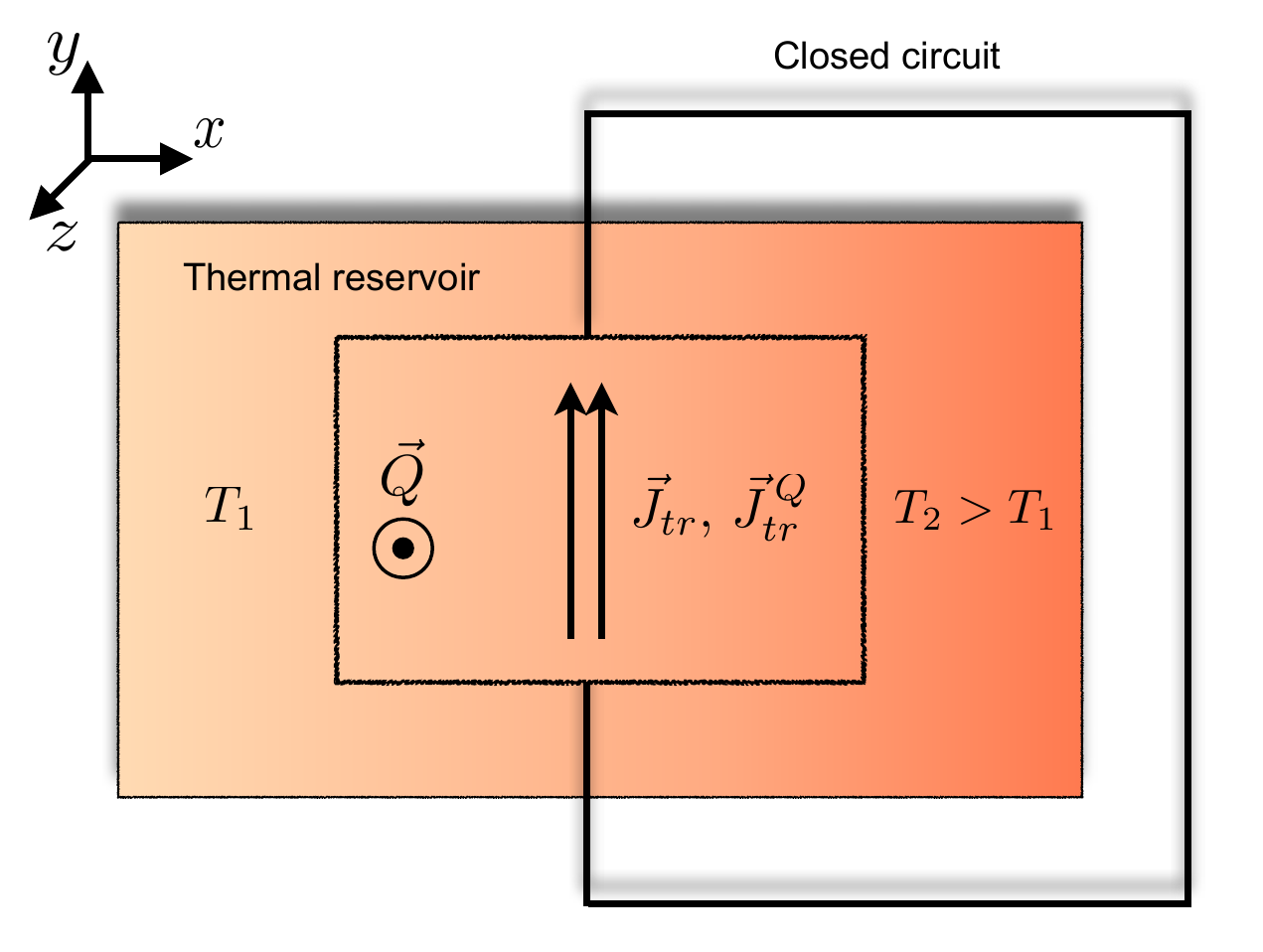}
\end{minipage}

\caption{Two possible experimental setups. (a) An open circuit consisting of a Weyl semimetal sample flanked by two thermal reservoirs at different temperatures. Electric and heat currents are not able to flow, and as a consequence an electric field, $\bm E$, and a gradient of temperature, $\bm\nabla T$, generate across the sample in the $x$ and $y$ directions, respectively. (b) A closed circuit made of a Weyl semimetal connected to current leads and surrounded by a thermal reservoir, in which a temperature gradient has been implemented. Electric and heat currents are free to flow.}
\label{fig. 1}
\end{figure*}

For a type-I Weyl semimetal with $|C|\ll v$,
from the Appendix \ref{appendix} and Eqs. (\ref{eq nernst conductivity}) and (\ref{eq thermal Hall conductivity}), in the limit of small tilt we find
\begin{gather}
\alpha_{xy} = -\frac{ek_B^2TC}{18\hbar^2 v^2},\label{eq alpha I}\\
K_{xy} = \frac{k_B^2T}{6\hbar} \bigg[Q -\frac{\mu}{2\hbar C}\bigg(\frac{v}{C}\ln\bigg|\frac{v+C}{v-C}\bigg|-2\bigg) \bigg]\nonumber\\
\approx\frac{k_B^2T}{6\hbar} \bigg[Q -\frac{\mu C}{3\hbar v^2} \bigg]\label{eq K I} .
\end{gather}
In our linearized model, the Nernst conductivity is linearly proportional to the tilt parameter $C$ and does not depend on the position of the Fermi level $\mu$, while the thermal Hall conductivity acquires an additional Fermi surface contribution compared to the untilted Weyl semimetal. 

For a type-II Weyl semimetal with $|C|\gg v$, deep in the type-II phase, we obtain
\begin{gather}
\alpha_{xy} = -\frac{ek_B^2T}{6\hbar^2}\frac{v}{|C| C}\left[-1+\ln\bigg|\frac{C^2 \Lambda}{v \mu}\bigg|\right],\label{eq alpha II}\\
K_{xy} = \frac{k_B^2T}{6\hbar}\frac{v}{|C|}\left[ Q - \frac{\mu}{\hbar C}\ln\bigg|\frac{C^2\Lambda}{v \mu}\bigg| \right]\label{eq K II}.
\end{gather}
Both expressions diverge logarithmically with the momentum cut-off $\Lambda$ along the $z$-axis, which as we previously mentioned is a signature of the presence of unbounded electron-hole pockets in the linearized model. 
In general, $\Lambda$ depends on $\mu$ and could be different for electron and hole pockets. 
For simplicity, and consistent with the fact that the linear model only gives results at the qualitative level, we fix in all Fermi surface integrals the cut-off $\Lambda$ to be independent of $\mu$. 
The divergent behavior of the Nernst conductivity in the limit $\mu\rightarrow0$ is a consequence of the logarithmic dependence of $\sigma_{xy}$ on $\mu$ for the type-II Weyl semimetal \cite{ZT16}.
We also observe that thermal Hall conductivity at $\mu=0$ is a factor $v/|C|$ smaller than in the type-I case. 

We briefly mention the case of inversion breaking tilt, with $C_+=C_-=C$. 
In this situation, in the linear model the contribution to the Hall conductivity from the Fermi surface cancels out between the two Weyl points, and consequently $\sigma_{xy}$ is independent of $\mu$. 
We then get
\begin{gather}
\alpha_{xy} = 0,\\
K_{xy} = \frac{k_B^2TQ}{6\hbar}\,\mathrm{min}\big(1,\frac{v}{|C|}\big),
\end{gather}
which is valid for general $C$.

Finally, a comment is in place regarding the possibility of having tilts orthogonal to the direction of separation of the Weyl nodes. 
The tilt always gives a Fermi surface contribution to the anomalous Hall conductivity in the plane perpendicular to the tilt direction \cite{ZT16}. Therefore, if in our setup we consider tilts in, say, the $x$ direction, instead of the $z$ direction, we obtain a finite anomalous Hall conductivity in the $y$-$z$ plane, in addition to the usual $\mu$ independent Hall conductivity, proportional to $Q$, in the $x$-$y$ plane. This leads to nonzero $\alpha_{yz},K_{yz}$ and $K_{xy}$, while $\alpha_{xy}=0$, with $K_{yz}$ being entirely a Fermi surface effect and $K_{xy}$ being $\mu$ independent and proportional to $Q$.

\section{Discussion of possible experimental setups}

In order to measure the Nernst and thermal Hall responses, one can think of two possible experimental setups.
The first one is the open circuit represented in Fig. \ref{fig. 1}(a), consisting of a Weyl semimetal sample attached to two thermal reservoirs at different temperatures, and disconnected from current leads. 
In this situation, the transport electric current in Eq. (\ref{eq transport 1}) vanishes and a voltage across the sample, given by Eq. (\ref{eq voltage}), is generated. For a clean sample, the longitudinal DC conductivity is zero, and from Eq. (\ref{eq Mott}) we get
\be
E_x=-\partial_x A_0=\sigma_{xy}^{-1}\,\alpha_{xy}.
\ee
That is, an electric field in the $x$ direction is generated, proportional to the Nernst conductivity. By measuring this voltage across the sample, $\alpha_{xy}$ can be inferred.

In this same open circuit setup, the thermal Hall effect in Eq. (\ref{eq thermal hall current}) tells us that heat is transported from one side of the sample to the other, perpendicularly to both the direction of the separation of the Weyl nodes and the temperature gradient generated by the thermal reservoirs. The heat current can be defined as the rate at which heat is transported
\be
J^Q_{\textrm{tr}}=\frac{k_B}{A}\frac{dT}{dt},
\ee
where $A$ is the cross-sectional area. Then, for setup of Fig. \ref{fig. 1}(a), the rate at which the temperature gradient increases between the upper and lower boundaries is
\be
\frac{k_B}{L_z}\frac{d\Delta T}{dt}=-K_{xy}(T_2-T_1),
\ee
with $L_z$ the thickness of the sample in the $z$ direction and $T_2-T_1$ the temperature difference between the two thermal reservoirs. This rate gives a measure of the thermal Hall conductivity $K_{xy}$.

The second possible setup is depicted in Fig. \ref{fig. 1}(b). Here we consider a closed circuit, where electric and heat currents can flow. The Nernst and thermal Hall conductivities calculated are transport coefficients, which means that they represent the net currents of the system, directly accessible by conventional transport experiments, without the need to probe the system locally. Therefore, the setup of Fig. \ref{fig. 1}(b) supports readily measurable Nernst and thermal Hall currents, given by Eqs. (\ref{eq nernst current},\ref{eq thermal hall current},\ref{eq alpha I},\ref{eq K I},\ref{eq alpha II},\ref{eq K II}).

\section{Conclussions}

We have studied the Nernst and thermal Hall responses of a minimal, linearized model for time-reversal-breaking tilted Weyl semimetals. 
In the situation of inversion-symmetric tilt, the anomalous Nernst conductivity is finite and the thermal Hall conductivity acquires Fermi surface contributions. 
This is in sharp contrast to the nontilted case, in which case the Nernst conductivity vanishes in the linear model and the thermal Hall conductivity does not depend on the Fermi level.  
On the other hand, for inversion-breaking tilt of equal size for the two chiralities, the Fermi surface contributions cancel out between the two Weyl cones. This leads to a vanishing Nernst conductivity and a chemical potential independent thermal Hall conductivity that decreases with increasing tilt in the type-II phase. 

We discussed two possible experimental setups, consisting in open and closed circuits. For the open circuit, the Nernst and thermal Hall conductivities can be inferred, respectively, by measuring an induced voltage across the sample and the rate at which a temperature gradient is generated between opposite boundaries. In the closed circuit case, the Nernst and thermal Hall currents are able to flow, and therefore could be directly measured. It is important to have in mind that the responses calculated are transport currents, which means that they represent the net current flow, made of both bulk and surface contributions, rather than the local total current at a given point of the sample. Therefore, the obtained Nernst and thermal Hall currents can be detected by conventional transport measurements.

\acknowledgments
We thank Maria A. H. Vozmediano, Alberto Cortijo, Karl Landsteiner and Adolfo G. Grushin for fruitful and enlightening discussions. This work was supported by the ERC Starting Grant No. 679722. A. A. Z. acknowledges support from the Swedish Research Council Grant No. 642-2013-7837, the Goran Gustafsson Foundation, and Academy of Finland.

\textbf{Note added}.-After submitting this work, the preprint~\onlinecite{ST17} appeared, which also studies the anomalous Nernst effect in type-II Weyl semimetals. Their results are compatible with ours.

\appendix

\section{Anomalous Hall conductivity for the tilted Weyl semimetal}
\label{appendix}
In this appendix we provide, for completeness and following Ref.~\onlinecite{ZT16}, expressions for the anomalous Hall conductivity. The anomalous Hall conductivity is given by the zero frequency and zero wave-vector limit of the current-current correlation function (we take $\hbar=1$)
\begin{eqnarray}\nonumber
\Pi_{ij}(\Omega,\mathbf{q}) &=& T\sum_{\omega_n}\sum_{s=\pm}\int \frac{d^3k}{(2\pi)^3} J_{i}^{(s)} G_{s}(i\omega_n,\mathbf{k})\\
&\times&J_{j}^{(s)} G_{s}(i\omega_n-i\Omega_m,\mathbf{k}-\mathbf{q})\bigg|_{i\Omega_m\rightarrow \Omega + i\delta},
\end{eqnarray}
where $i,j=\{x,y,z\}$, $T$ is the temperature (the Boltzmann constant is set to unity), and $\omega_n$, $\Omega_m$ are the fermionic and bosonic Matsubara frequencies, as
\begin{eqnarray}
\sigma_{xy} = -\lim_{\Omega\rightarrow 0}\frac{\Pi_{xy}(\Omega,0)}{i\Omega}.
\end{eqnarray}
The one-particle Green functions have the following form
\begin{equation}
G_{s}(i\omega_n,\mathbf{k}) = \frac{1}{2}\sum_{t=\pm1} \frac{1-st\bm\sigma \cdot \frac{\mathbf{k}-sQ\mathbf{e}_z}{|\mathbf{k}-sQ\mathbf{e}_z|}}{i\omega_n+\mu -C_{s}(k_z-sQ) +t v |\mathbf{k}-sQ\mathbf{e}_z|},
\end{equation}
where $\mu$ is the chemical potential.
The current operators are defined as follows
\begin{equation}
J_{i}^{(s)} = e  \left(C_{s}\delta_{iz} +sv \sigma_i \right),
\end{equation}
where $\delta_{ij}$ is the Kronecker delta. Performing the summation over the fermionic frequencies and taking the trace over Pauli $\sigma$-matrices we obtain 
\begin{equation}
\Pi_{xy}(\Omega,0) = \Pi_{xy}^{(+)}(\Omega,0)+\Pi_{xy}^{(-)}(\Omega,0),
\end{equation}
where we separated the contributions from the two Weyl cones
\begin{gather}
\Pi_{xy}^{(s)}(\Omega,0)= \Pi_0^{(s)}(\Omega,0)+ \Pi_{\mathrm{FS}}^{(s)}(\Omega,0),\\
\nonumber
\Pi_{0}^{(s)}(\Omega,0)= se^2 \int_{-\Lambda_0 -sQ}^{\Lambda_0 - sQ} \frac{dk_z}{2\pi}\int_0^{\infty} \frac{k_{\perp}dk_{\perp}}{2\pi} \frac{2v^2 \Omega_m}{\Omega_m^2+4v^2k^2}\\
\times
\frac{k_z}{k}\bigg|_{i\Omega_m\rightarrow \Omega + i\delta},\\
\nonumber
\Pi_{\mathrm{FS}}^{(s)}(\Omega,0)= se^2 \int_{-\Lambda -sQ}^{\Lambda - sQ} \frac{dk_z}{2\pi}\int_0^{\infty} \frac{k_{\perp}dk_{\perp}}{2\pi} \frac{2v^2 \Omega_m}{\Omega_m^2+4v^2k^2}\\
\times
\frac{k_z}{k} \bigg \{ f(C_{s}k_z+vk)-  f(C_{s}k_z-vk)-1
 \bigg\}\bigg|_{i\Omega_m\rightarrow \Omega + i\delta}.
\end{gather}
$\Pi_{0}$ is the vacuum part, this is the contribution at $\mu=0$, and $\Pi_{\mathrm{FS}}$ is the contribution of the states at the Fermi surface. 
The function $f(E) = (e^{(E-\mu)/T}+1)^{-1}$ gives the Fermi distribution and $k = \sqrt{k_z^2+k_{\perp}^2}$. The cut-off $\Lambda_0$ is just included for the correct definition of the $k_z$ integral, but the vacuum contribution to the Hall conductivity is well known to be cut-off independent. However, we have introduced another cut-off $\Lambda$  for the Fermi surface contribution. This has a physical meaning, as explained in the main text, and is necessary in order to make the Fermi surface contribution finite in the type-II regime.

We obtain the anomalous Hall conductivity in the limit of zero frequency $\sigma_{xy} = \sigma_{xy}^{(+)}+\sigma_{xy}^{(-)}$, where
\begin{gather}
\sigma_{xy}^{(s)}= \sigma_0^{(s)}+ \sigma_{\mathrm{FS}}^{(s)},\\
\sigma_0^{(s)}= e^2 \int_{-\Lambda_0 -sQ}^{\Lambda_0 - sQ} \frac{dk_z}{2\pi}\int_0^{\infty} \frac{k_{\perp}dk_{\perp}}{2\pi}\frac{sk_z}{2k^3},\\
\nonumber
\sigma_{\mathrm{FS}}^{(s)}= e^2 \int_{-\Lambda -sQ}^{\Lambda - sQ} \frac{dk_z}{2\pi}\int_0^{\infty} \frac{k_{\perp}dk_{\perp}}{2\pi}\\
\times
\frac{sk_z}{2k^3} \bigg[f(C_{s}k_z+vk)-f(C_{s}k_z-vk)-1\bigg],
\end{gather}
in which $sk_z/2k^3$ is the $z$-component of the Berry curvature of the Weyl cone with chirality $s$. Here we note that the tilt parameter does not modify the Berry curvature. Thus the tilt enters only in the distribution function.

Taking the zero temperature limit, $T\rightarrow 0$, and performing the integration over momentum $k_{\perp}$, we obtain the contribution to the anomalous Hall conductivity in the form
\begin{gather}
\sigma_0^{(s)}= -\frac{se^2}{8\pi^2}\int_{-\Lambda_0 -s Q}^{\Lambda_0 -s Q} dk_z\,\mathrm{sign}(k_z),\\
\nonumber
\sigma_{\mathrm{FS}}^{(s)}=-\frac{se^2}{8\pi^2}\int_{-\Lambda -s Q}^{\Lambda -s Q} dk_z\left[\mathrm{sign}(k_z)-\frac{vk_z}{|C_{s}k_z-\mu |}\right]\\
\times\left[(\Theta\left(v^2 k_z^2 - (C_{s}k_z -\mu)^2\right)-1\right],
\label{eq:sigmaxy12kz}
\end{gather}
where $\Theta(x)$ is the Heaviside function. The expression Eq. (\ref{eq:sigmaxy12kz}) is valid for any $C_{+}$ and $C_-$; however, it can be simplified further by taking into account the condition $C_+=-C_-\equiv C$. Summing the contribution of the two Weyl cones we obtain
\begin{gather}
\sigma_{xy}=\sigma_0+ \sigma_{\mathrm{FS}}\\
\sigma_0= -\frac{e^2}{4\pi^2}\int_{-\Lambda_0 - Q}^{\Lambda_0 - Q} dk_z\,\mathrm{sign}(k_z),\\
\nonumber
\sigma_{\mathrm{FS}}=-\frac{e^2}{4\pi^2}\int_{-\Lambda - Q}^{\Lambda - Q} dk_z\left[\mathrm{sign}(k_z)-\frac{vk_z}{|Ck_z-\mu |}\right]\\
\times\left[(\Theta\left(v^2 k_z^2 - (Ck_z -\mu)^2\right)-1\right],
\end{gather}

By performing the integrals we can readily evaluate $K_{xy}$, and by taking the derivative of the result we obtain $\alpha_{xy}$. There are a few limiting cases in which analytic expressions can be obtained, as given in the main text.


%

\end{document}